\documentclass[aps,pre,groupedaddress,superscriptaddress,twocolumn]{revtex4-1}

\usepackage{graphicx}

\def\ER{Erd\H{o}s-R\'enyi }

\begin{document}

\title{Targeted Damage to Interdependent Networks}

\author{G. J. Baxter}
\email[]{gjbaxter@ua.pt}
\author{G. Tim{\'a}r}
\author{J. F. F. Mendes}
\affiliation{Department of Physics \& I3N, University of Aveiro, Campus
  Universit\'ario de Santiago, 3810-193 Aveiro, Portugal}

\date{\today}

\begin{abstract}
Complex systems consisting of interdependent subsystems may be represented by multi-layer networks, with interdependency links between layers. The giant mutually connected component (GMCC) of such an interdependent (or multiplex) network collapses with a discontinuous hybrid transition under random damage to the network. If the nodes to be damaged are selected in a targeted way, the collapse of the GMCC may occur significantly sooner. Understanding the limits of the resilience of such systems to targeted attacks is therefore an essential problem. Finding the minimal damage set which destroys the largest mutually connected component of a given interdependent network is a computationally prohibitive simultaneous optimization problem. We introduce a simple heuristic strategy---Effective Multiplex Degree---for targeted attack on interdependent networks that 
leverages the indirect damage inherent in multiplex networks to achieve 
 a damage set smaller than that found by any existing non-computationally intensive algorithm. We show that the intuition from single layer networks that decycling (damage of the $2$-core) is the most effective way to destroy the giant component, does not carry over to interdependent networks, and in fact such approaches are worse than simply removing the highest degree nodes.
\end{abstract}

\maketitle


\section{Introduction}

Interdependent networked systems are particularly vulnerable to damage, as damaged sites in one sub-system cause failures in another, which may in turn propagate damage back to the first \cite{buldyrev2010catastrophic}. Under sufficient damage, avalanches of propagating failures may eventually lead to the discontinuous collapse of the whole system \cite{baxter2012avalanche}.
Many essential natural, technological and social systems \cite{pocock2012robustness,rinaldi2001identifying,szell2010multirelational} consist of fully or partially interdependent subsystems, allowing them to be represented by such interdependent or multiplex networks. Understanding the vulnerability of such systems to damage and cascading collapse is therefore of paramount importance.

If the nodes damaged are selected non-randomly, that is, the damage is targeted, the amount of damage required for collapse may be dramatically reduced. In this paper,
we consider the problem of finding the best approximation to the minimal damage set: a set of nodes whose removal leads to the collapse of the largest mutually connected component (LMCC). We show that the characteristics of highly effective strategies on interdependent networks are different from those on single layer networks. 
We introduce a new heuristic targeted attack strategy which exploits the properties of interdependent or multiplex percolation to effectively destroy the LMCC more quickly than any previous non-computationally intensive method, with an efficacy approaching that of simulated annealing, yet with a computation time for node ranking that is linear in system size.

The quest to identify the minimal set of removals required to destroy the giant connected component in a single network has recently received renewed attention. 
The \texttt{MinSum} algorithm \cite{braunstein2016network} combines a message-passing approach with a statistical mechanics formulation, achieving results which outperform even stochastic search methods such as simulated annealing. The approach is essentially similar to the \texttt{MaxSum} \cite{altarelli2013optimizing} method proposed for threshold problems. While much more computationally efficient than simulated annealing, \texttt{MinSum} is nevertheless rather complex and opaque. Several more simple heuristic methods have also been proposed, such as collective influence (\texttt{CI}) 
\cite{morone2015influence,morone2016collective}, and \texttt{CoreHD}, a direct attack on the 2-core of the largest component \cite{zdeborova2016fast}. Despite its simplicity, the \texttt{CoreHD} algorithm performs essentially as well as \texttt{MinSum}.
These methods exploit the observation that the destruction of the largest connected component is in grand part a matter of decycling the network, i.e. removing all loops \cite{braunstein2016network}.
Once this is done, the remaining tree can be broken down in very few steps. All loops in a network are contained within its $2$-core, the subnetwork in which each member has at least $2$ neighbours which are also members. Hence attacking the $2$-core of a network is an effective strategy for dismantling a network.

\begin{figure}[htb]
\includegraphics[width=0.98\columnwidth]{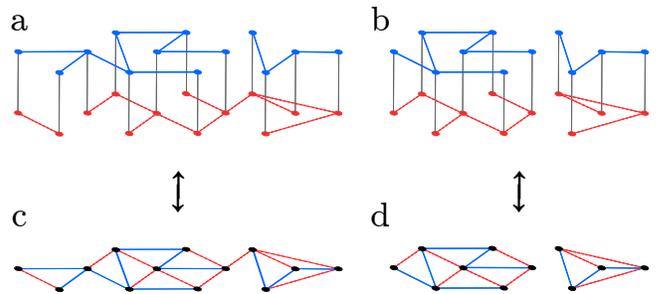}
\caption{(Color online.) (a) An example of a (fully) interdependent two-layer network. (b) The mutually connected components present in this network. (c) The multiplex representation of the same network, with (d) the mutually connected components.}\label{fig_diagram}
\end{figure}

Here we consider the equivalent problem for multiplex and interdependent networks, where similar progress has not yet been achieved.
An interdependent network consists of two or more network layers, with nodes in different layers connected by interdependency links, as shown in Fig. \ref{fig_diagram} (a). If a node fails in (is removed from) one layer, its interdependent partner in another layer also fails. In the multiplex network representation,  interdependent nodes are merged into a single node, and connections in different layers are considered different types of edge (represented by different colors) \cite{son2012percolation}, Fig. \ref{fig_diagram} (c). Mathematically the two representations are equivalent.

The generalization of a connected cluster to multiplex or interdependent networks is the mutually connected component (MCC). A node belongs to a MCC if at least one of its neighbors in the same layer belongs to the same MCC, and each of its interdependent partners in other layers also belong to the same MCC \cite{buldyrev2010catastrophic}, as shown in Fig. \ref{fig_diagram} (b). In the multiplex representation  (for fully interdependent networks) this  corresponds to the rule that all pairs of nodes within an MCC must be connected by at least one path of each edge type (color) \cite{son2012percolation, baxter2012avalanche,baxter2016unified},  Fig. \ref{fig_diagram} (d).
In the thermodynamic limit, the collapse of a giant mutually connected component (GMCC) under random damage occurs with a discontinuous hybrid phase transition \cite{buldyrev2010catastrophic, baxter2012avalanche}. The fraction of vertices that must be randomly removed to provoke the collapse of the system  depends on the structure of the two layers. 

Huang {\em et al.} examined the effect of removing nodes preferentially according to their degree \cite{huang2011robustness}, and the results were extended to partially interdependent networks in \cite{dong2012percolation}. In \cite{osat2017optimal} the authors consider various existing strategies, finding the best of them is the limiting case of that in \cite{huang2011robustness}: remove the nodes with the highest degree sum or product. 
Similar cascade phenomena occur in single-layer networks under threshold criteria, such as the $k$-core or bootstrap percolation \cite{pittel1996sudden,balogh2007bootstrap,baxter2015critical,baxter2011heterogeneous}. The \texttt{MaxSum} algorithm \cite{altarelli2013optimizing} provides a message-passing based method to find good solutions to the minimal seeding set problem in such threshold models.
Such methods could in principal be applied to the $k$-core process, but cannot be directly applied to the multiplex percolation problem, because a node's membership of the LMCC cannot be computed directly from the state of its neighbors, but instead requires an exploration of the entire cluster \cite{baxter2016unified}. In other words there is no local pruning rule that allows one to identify the LMCC.

\section{Targetted Damage Strategies}

We consider a configuration model multi-layer network with $M$ layers.
Each node $i$, $i\in\{1,2,...N\}$ in layer $l$ is interdependent with a single node (for convenience also labeled $i$) in each of the other layers, such that removal of $i$ from layer $l$ implies the failure (removal) of all counterpart nodes $i$ in the other layers \cite{buldyrev2010catastrophic}. Alternatively, by considering all the interdependent nodes $i$ as a single node spanning all layers, the system can be viewed as a multiplex network, defined by the joint degree distribution $P(q^{(1)},q^{(2)},...,q^{(M)}) \equiv P(\bf{q})$,  such that node $i$ has degree $q_i^{(l)}$ in layer $l$. These two formulations, multiplex and interdependent networks, are mathematically equivalent \cite{son2012percolation}. In Sect. \ref{partial}, below, we extend our analysis to partially interdependent networks, in which some nodes do not have interdependent partners.

In the limit $N\to \infty$, the GMCC is extensive if it contains a finite fraction of the nodes in the multiplex. In large finite networks, we consider the largest mutually connected component (LMCC) to be extensive if its size $S$ exceeds $\sqrt{N}$. 
Starting from a multiplex in which $S > \sqrt{N}$, our aim, then, is to select the smallest possible damage set $s$ that destroys the LMCC.

A given attack strategy ranks nodes according to some metric, and nodes are removed in order starting from the highest ranked node, until the LMCC collapses. The fraction $p_c \equiv |s|/N$ of removed nodes is the basis of comparison between different methods.
An adaptive strategy may be used, with ranks being recalculated after the network reaches equilibrium following each removal.
This also avoids the possibility of redundancies caused by removing nodes that may have been removed anyway by avalanches.
Only nodes belonging to the LMCC at the current step are considered for removal (including for random node removal).
Note that the final result depends only on the set of nodes removed, not on the order of removal.

\begin{figure}[htb]
\includegraphics[width=0.8\columnwidth]{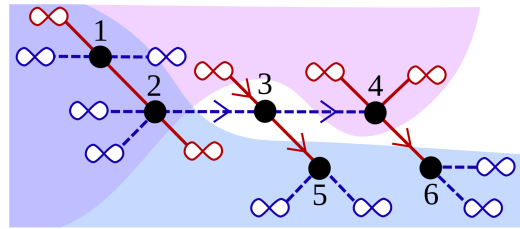}	
\caption{(Color online.) A small grouping of nodes within the GMCC (LMCC) of a two layer multiplex. Infinity symbols indicate a connection leading to an infinite GMCC subtree (or the LMCC $2$-core in a finite system) in a given layer. Two or more such connections correspond to membership of the GMCC $2$-core of that layer. Nodes $1$, $2$, $5$, and $6$ belong to the $2$-core within the GMCC in the layer with dashed (blue) edges. Nodes $1$, $2$, and $4$ belong to the $2$-core of the solid (red) edged layer. Since nodes $1$ and $2$ belong to the $2$-core of the GMCC in both layers they cannot be removed in an avalanche. Nodes $3$-$6$ are outside one or both $2$-cores and are therefore in a critical state, potentially being removed in an avalanche of propagating damage. For example, removal of node $2$ will lead to the removal of all four of these nodes. Damage propagates along critical edges in the direction marked by arrows.}\label{fig_node_types}
\end{figure}

\begin{figure}[htb]
\includegraphics[width=0.95\columnwidth]{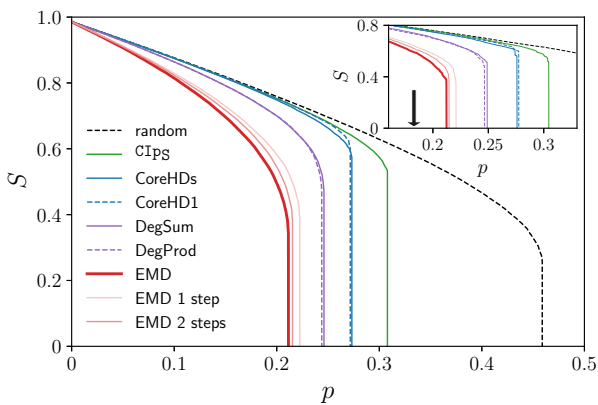}	
\caption{(Color online.) Progression of the relative size, $S$, of the LMCC with a fraction $p$ of nodes removed, under different targeted attack strategies for a duplex network of two uncorrelated \ER layers with mean degree $\mu=5$ and $N=10^5$. From right to left: random damage (dashed, black), collective influence $\texttt{CIp}_\texttt{S}$(green), \texttt{CoreHDs} (solid blue), \texttt{CoreHD1} (dashed blue),highest degree sum \texttt{DegSum} (purple) and product \texttt{DegProd} (dashed purple), Effective Multiplex Degree \texttt{EMD} (red). Also shown are the approximate effective multiplex degree results obtained using only one and two iterations of Eq. (\ref{w_matrix}).
Inset: Results for $N=10^4$, allowing comparison with simulated annealing, marked by the arrow.}\label{fig_strategy_comparison}
\end{figure}

In interdependent networks, 
critical nodes are nodes that depend on a single connection in one (or more) layers in order to remain within the LMCC.
Damage propagates through clusters of such critical nodes, as the removal of one node removes the support from neighboring critical nodes, and so on, as illustrated in Fig. \ref{fig_node_types}. The avalanches of damage diverge in size approaching the critical point, leading to the discontinuous collapse of the GMCC.
The $2$-cores in each layer within the LMCC play an important role. 
Nodes within the LMCC $2$-core in a given layer are not critical with respect to that layer, so that nodes belonging to the $2$-core in all layers are non-critical, while 
nodes failing to belong to one or more of these $2$-cores are in a critical state and may be removed by avalanches, see Fig. \ref{fig_node_types}. 
One might therefore imagine that, as in single-layer networks, destruction of $2$-cores might be an effective strategy in interdependent networks.
We therefore generalized the leading single-layer strategies which target $2$-cores, to see if they would be effective also in multiplex networks.
Typical results are shown in Fig. \ref{fig_strategy_comparison}. 
We found that the intuitions gained from studying single layer networks do not carry over to multiplex networks. 

The collective influence (\texttt{CI}) strategy may be adapted to multiplex networks by selecting nodes according to the sum or the product of their scores in each layer. 
We use the collective influence propagation \texttt{CIp} algorithm \cite{morone2016collective}, and results shown are for the sum $\texttt{CIp}_\texttt{S}$.
Similarly, the \texttt{CoreHD} method may be generalized by removing the nodes with the highest sum of degrees within the union of the 2-cores of all layers (\texttt{CoreHDs}) or by removing node with the highest degree within the $2$-core of the layer with the (currently) lowest mean degree (\texttt{CoreHD1}). See Appendix \ref{strategies} for more details.
We found, however, that simply removing the node with the highest degree sum (\texttt{DegSum}) or product (\texttt{DegProd}) is more effective than any of these strategies, 
meaning that focusing on the $2$-cores actually diminishes attack effectiveness.
This further underlines the fundamentally different nature of the multiplex percolation transition.

\section{Effective Multiplex Degree}

Here we introduce a new strategy, Effective Multiplex Degree (\texttt{EMD}) which gives a damage set significantly smaller than that found by any of the methods described above. 
The \texttt{EMD} algorithm leverages inhomogeneities between layer degrees and in the neighborhood of a node to maximize the effect of node removal.

We seek to ascribe weights $w_i$ which represent the impact on the LMCC of removing node $i$. Consider a connection between nodes $i$ and $j$ in layer $l$. If $i$ is removed, $j$ loses an edge in layer $l$.
 If node $j$ loses all of its $q_j^{(l)}$ connections within layer $l$, it is removed from the LMCC. A node is most vulnerable in the layer in which it has least connections. The smaller is $q_j^{(l)}$, the bigger is the impact of losing one of those connections. Suppose that node $j$ has weight $w_j$, then we ascribe a value $w_j/q_j^{(l)}$ to the impact of removing the edge from $i$ to $j$ in layer $l$. The weight of a node $i$ is then the sum of all the weights ascribed to the edges emanating from it, a measure of the impact of removing $i$:
\begin{equation}\label{w_i}
w_i = \sum_{l=1}^M \sum_{j \in \mathcal{N}_i^{(l)}} \frac{1}{M_j}\frac{w_j}{q_j^{(l)}},
\end{equation}
where $\mathcal{N}_i^{(l)}$ is the set of neighbors of node $i$ in layer $l$,
and $M_j$ is the number of layers in which node $j$ participates. This naturally accounts for the case of partial interdependence, see Sect. \ref{partial} below.
The \texttt{EMD} weights are the self-consistent solutions of this set of equation. Larger weights are found in nodes with more connections, and with connections to other highly weighted nodes in the layer in which they are weakest.

Eq. (\ref{w_i}) may be written more compactly as the matrix equation
\begin{equation}\label{w_matrix}
\bf{w} = \bf{Rw}
\end{equation}
with 
\begin{equation}\label{R}
R_{ij} = \sum_{l=1}^M \frac{a_{ij}^{(l)}}{M_jq_j^{(l)}}
\end{equation}
where $a_{ij}^{(l)}$ are the elements of the adjacency matrix in layer $l$.
The matrix $\mathbf{R}$ is a left stochastic matrix. Since the weights $w_i$ are all positive, by the Perron-Frobenius theorem, the vector $\mathbf{w}$ is the leading right eigenvector of $\mathbf{R}$, corresponding to the largest eigenvalue $1$.

The weight $w_i$ is equal to the equilibrium probability to find a random walker at node $i$, when the random walker at each step chooses uniformly between layers, and selects an edge uniformly within that layer to follow.
In \cite{dedomenico2013random} a type of random walk is introduced which is a classical random walk within the layers and at each node the walker is allowed to randomly switch to the multiplex counterpart node in a negligible time. This process is equivalent to the one described by our matrix $\bf{R}$ in a fully interdependent network of an arbitrary number of layers. Note, however, that the random walk corresponding to matrix $\mathbf{R}$ is defined for an arbitrary network of networks, in which any node can have an arbitrary number of interdependency neighbours in other layers.
 
In \cite{iacovacci2016functional} a node centrality index, called Functional Multiplex PageRank, is introduced. This corresponds to a random walker on a general directed multiplex network with an arbitrary combination of overlapping links. Although the random walk transition probability, from a node to its neighbour, may depend on the type of link overlap, it is essentially determined by an ``aggregated degree" of each node. The major advantage of EMD over this centrality definition is that it incorporates the degree-heterogeneity of nodes over different layers. This degree-heterogeneity is essentially important for the robustness of multiplex networks or the dynamical processes running on top of them. 
A number of other variations of random walks have been defined for interdependent networks \cite{sole-ribalta2016random,kuncheva2015community} which treat interdependency links as qualitatively the same - from the point of view of the random walker - as intralayer connectivity links. 

\begin{figure}[htb]
\includegraphics[width=0.65\columnwidth]{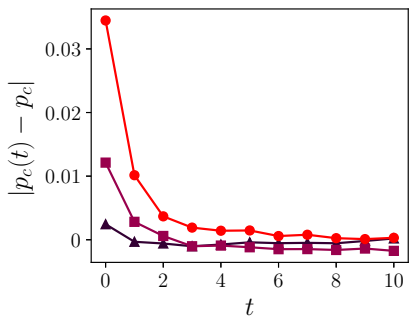}	
\caption{Absolute difference between damaging fraction $p_c(t)$ obtained by limiting to $k$ iterations of Eq. (\ref{w_matrix}) and final value $p_c$, for two fully interdependent \ER layers with mean degree $\mu=5$ and $N=10^5$ nodes, averaged over 10 realizations (circles, red). Initial value $t=0$ corresponds to highest degree sum strategy. Also shown are progressions for partially interdependent networks, $f=0.6$ (squares) and $f=0.3$ (triangles).\label{fig_w_convergence}}
\end{figure}

A simplified, non self-referential weight may be obtained by using the degree sum $Q_j = \sum_{l=1}^M q_j^{(l)}$
in place of $w_j$ in Eq. (\ref{w_i}):
\begin{equation}\label{w_i_simple}
\tilde{w}_i = \sum_{l=1}^M \sum_{j \in \mathcal{N}_i^{(l)}} \frac{1}{M_j}\frac{Q_j}{q_j^{(l)}}.
\end{equation}
This already gives significant improvement over \texttt{DegSum} and \texttt{DegProd}, see results \texttt{EMD1} in Fig. \ref{fig_strategy_comparison}. 
The convergence to $\bf{w}$ can be examined by considering the sequence of weights obtained by iterating Eq. (\ref{w_matrix}). We use as initial values the degree sum, $w_i^{(0)} = Q_i$, and define ${w_i}^{(t+1)} = \sum_j R_{ij}w_j^{(t)}$, so that $w_i^{(1)} = \tilde{w}_i$, and $w_i = w_i^{(t\to\infty)}$.
In Fig. \ref{fig_w_convergence} we plot the collapse threshold $p_c$ obtained by restricting the number of iterations of Eq. (\ref{w_matrix}).
We see that the result converges rapidly after only a very few iterations.
For a fixed number $t$ of iterations, the calculation of ${\bf{w}}^{(t)}$ is linear in system size. See Appendix \ref{CONV} for a more detailed discussion.

\section{Results}

To illustrate the results obtained by our method, in Fig. \ref{fig_strategy_comparison} we consider a multiplex consisting of two \ER layers with equal mean degree, and plot the relative size $S$ of the largest mutually connected component as a function of the fraction $p$ of nodes removed, starting with the largest weight $\max\{w_i\}$, and recalculating the weights after each removal. The LMCC collapses at a fraction $p_c$ that is significantly lower than for any of the other methods described. The result approaches that found by simulated annealing on a relatively small network ($N=10^4$). The computational budget required to perform simulated annealing grows rapidly with $N$, and already for $N=10^5$, our method outperforms the simulated annealing results achievable in a reasonable computing time (see Appendix \ref{SA}).
Furthermore, the surviving LMCC size $S$ falls more rapidly than in other methods, being smaller at every value of $p$.

\begin{figure}[h!tb]
\includegraphics[width=0.48\columnwidth]{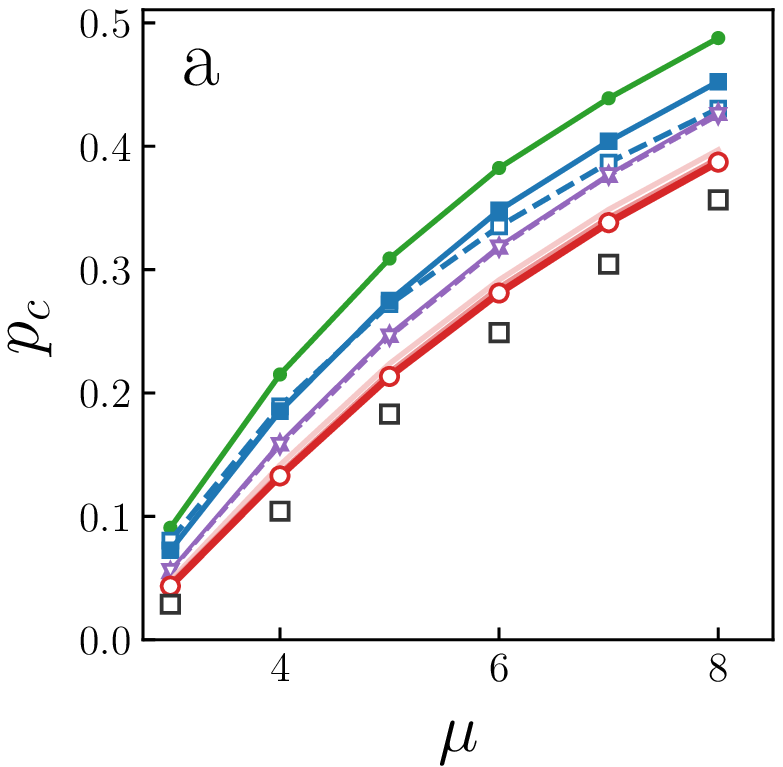}	
\includegraphics[width=0.48\columnwidth]{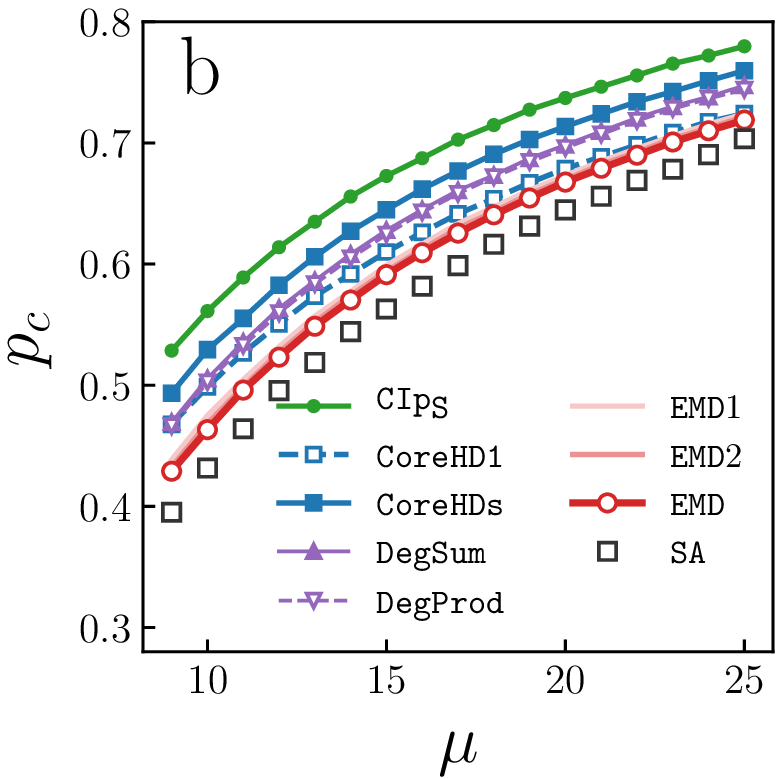}	
\caption{(Color online.) 
Results for the relative number of nodes removed in destruction of the LMCC for 
two \ER layers, both with varying mean degree $\mu$. (a) Values of $\mu$ from 3 to 8 (b) $\mu$ from 9 to 25. All networks consist of $N = 10000$ nodes. The networks are fully interdependent and the two layers are uncorrelated. All results were averaged over $10$ realizations, except for Simulated Annealing (\texttt{SA}), which was a single realization.
\label{figs_vs_mu}}
\end{figure}

\begin{figure}[htb]
\includegraphics[width=0.48\columnwidth]{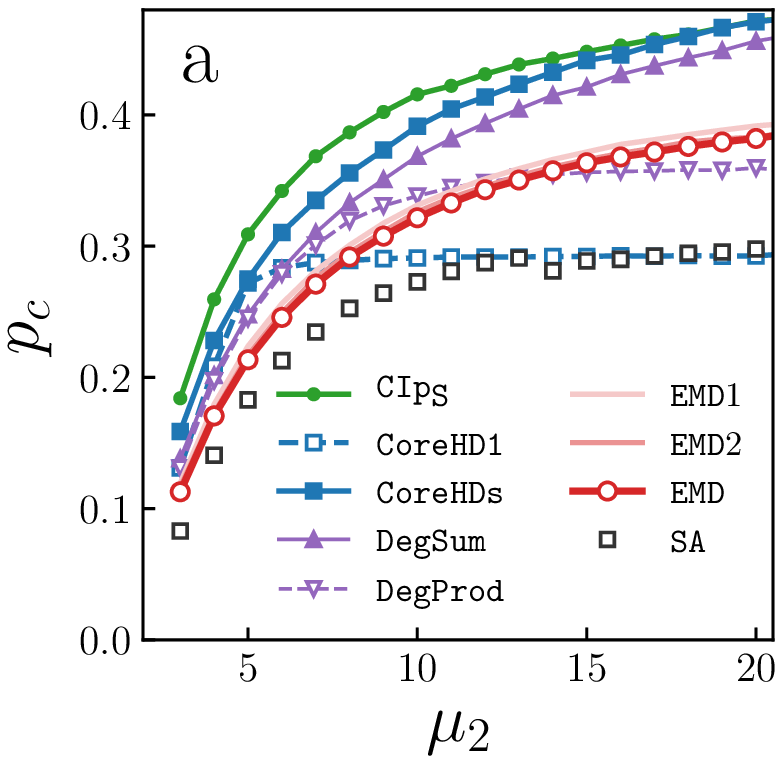}	
\includegraphics[width=0.48\columnwidth]{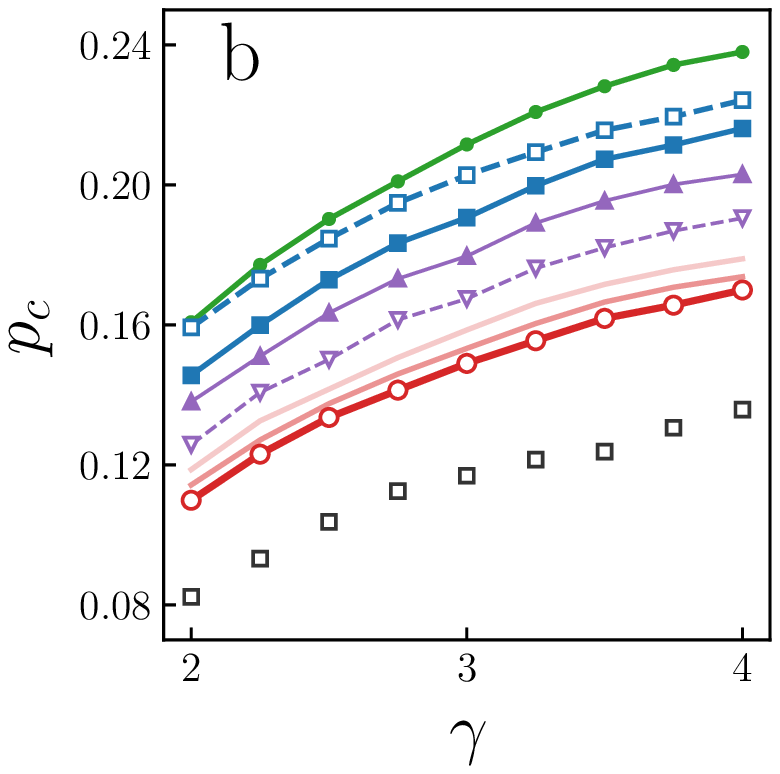}	
\caption{Collapse threshold $p_c$ for different attack strategies for 
(a) two uncorrelated \ER layers, with different mean degrees. One layer has fixed mean degree $\mu_1=5$ while the other has mean degree $\mu_2$. \label{fig_different} and
(b) two power-law degree distributed layers, $P(q)\propto q^{-\gamma}$ as a function of powerlaw exponent $\gamma$.
\label{fig_sf_comparison}
Each layer has $N=10,000$, 10 realizations, except simulated annealing, 1 realization.
}
\end{figure}

\begin{figure}[h!tb]
\includegraphics[width=0.48\columnwidth]{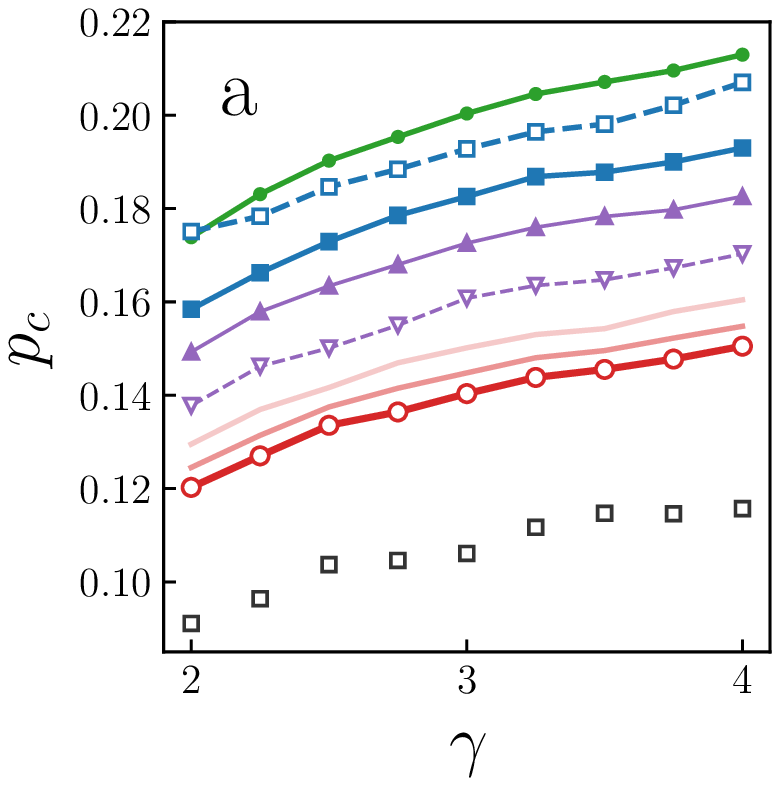}	
\includegraphics[width=0.48\columnwidth]{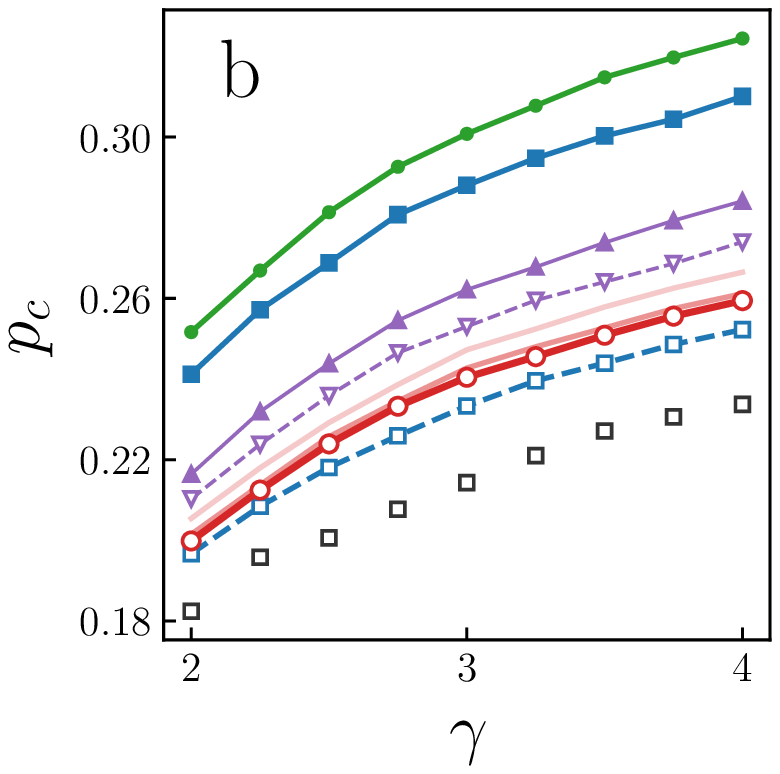}	
\caption{(Color online.) 
Results for the relative number of nodes removed in destruction of the LMCC for 
(a) two scale-free layers, both with mean degree $10$, one with exponent $\gamma_1 = 2.5$, the other with varying exponent $\gamma_2$ and (b) An \ER layer and a scale-free layer both with mean degree $10$, and varying degree distribution exponent $\gamma$. In this case the $\texttt{CoreHD1}$ algorithm was modified to always attack the scale-free layer.\label{figs_diverse}}
\end{figure}

To understand the range of effectiveness of the method, we investigated several combinations of \ER and uncorrelated scale-free layers to compare the performance of our \texttt{EMD} algorithm with other attack strategies, in different scenarios.
In Fig. \ref{figs_vs_mu} we illustrate the dependence on network density. We plot the results for two \ER layers with equal mean degree $\mu$. As $\mu$ increases the performance ranking of the different attack methods remains the same, with the exception of \texttt{CoreHD1}, which begins to become relatively more effective. With very few vulnerable nodes in dense networks, reducing the density of one of the layers starts to become a more effective strategy. Our \texttt{EMD} algorithm remains the most effective across all the values of $\mu$ tested. 

If one layer (of a two layer network) is much more dense than the other, the membership in the LMCC of a pair of interdependent nodes is effectively decided by the node in the less dense layer, and the network behaves as a single layer network. In this case, single layer strategies may be more effective than multiplex ones. To explore this possibility, we considered a duplex network with the mean degree of one layer, $\mu_1$, fixed, and varied the mean degree $\mu_2$ of the second layer, Fig. \ref{fig_different} (a). For large $\mu_2$, the best single layer strategy, \texttt{CoreHD1} becomes the leading strategy, confirming the fact that beyond a certain density in layer $2$, the destruction of the LMCC is equivalent to the destruction of the largest connected component in layer $1$.

To examine the effect of heterogeneity in network degree, we considered two scale-free layers. When the two layers have the same powerlaw distribution, with exponent $\gamma$, varying $\gamma$ has no effect on the ranking of the algorithms, with \texttt{EMD} always outperforming the other methods, Fig. \ref{fig_sf_comparison} (b). Similarly, if we fix the exponent for one layer, and vary it for the other layer, again the relative effectiveness of the different methods is unchanged, Fig. \ref{figs_diverse}(a).

Finally we considered what happens when one layer is of a significantly different character than the other, by pairing an \ER layer with a scale-free layer, each with the same mean degree, Fig. \ref{figs_diverse}(b). \texttt{DegProd} improved relative to the other methods. We modified the \texttt{CoreHD1} method to attack only the scale-free layer, as highly heterogeneous networks are more fragile under targetted attack, and found that this algorithm outperformed even \texttt{EMD}, reinforcing our results for layers of different densities, that when one layer is significantly more fragile than the other, single layer strategies are more effective.

\section{Partial Interdependence}\label{partial}

\begin{figure}[htb]
\includegraphics[width=0.6\columnwidth]{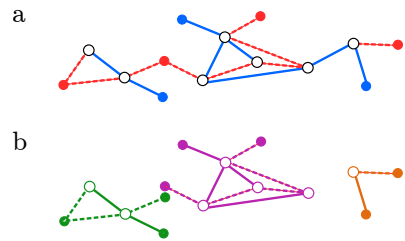}
\caption{(Color online.) (a) The multiplex representation of a partially interdependent two-layer network. Interdependent nodes are shown as open circles, non interdependent nodes as closed circles. In (b) we show the mutually connected components present in this network. Note that non interdependent nodes may belong to more than one MCC.}\label{fig_diagram_partial}
\end{figure}

The condition that every node is interdependent with nodes in other layers is a strong one. This may be relaxed to give a more realistic model  by considering partially interdependent two-layer networks \cite{parshani2010interdependent}, parameterized by the fraction of nodes $f$ within a given layer which are interdependent. When $f=1$, the network is fully interdependent. 
When $f=0$ the layers are not interdependent at all, and the system behaves as a pair of single-layer networks. In the multiplex representation, the fraction of interdependent nodes in the entire system is $\tilde{f} = f/(2-f)$. A simple example of such a network is illustrated in Fig. \ref{fig_diagram_partial} (a).
Under random damage, the collapse of the GMCC transitions from a hybrid discontinuous transition to a second order percolation transition for sufficiently small $f$ \cite{parshani2010interdependent}.

\begin{figure}[htb]
\includegraphics[width=0.98\columnwidth]{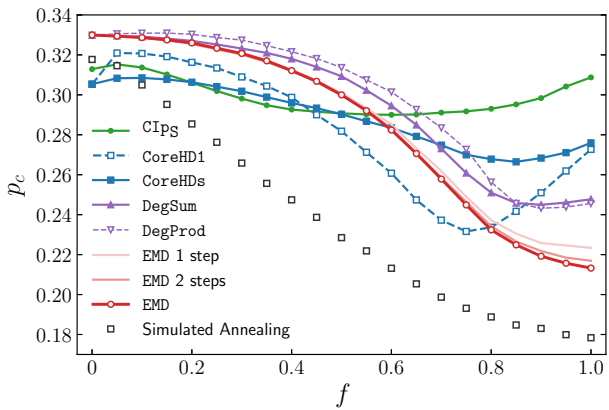}	
\caption{Comparison of collapse threshold $p_c$ as a function of interdependent node fraction $f$ for different targeted attack strategies for a duplex network of two uncorrelated \ER layers with mean degree $\mu=5$ and $N=10,000$, averaged over 50 realizations.
Simulated annealing results were averaged over five realization.
\label{fig_partial}}
\end{figure}

The definition of mutually connected components in this case is not as obvious as for fully interdependent networks. Usually a recursive definition of the giant mutually connected component (GMCC) is used, whereby a node belongs to the GMCC
 if it has at least one link to the GMCC 
in its own layer, and all of the nodes with which it is interdependent also belong to the GMCC. In terms of multiplex networks, this corresponds to the condition that a node has at least one link to the GMCC
 in each of the layers in which it participates. This definition applies both to fully and partially interdependent networks. However, we are not aware of a strict definition of \textit{finite} mutually connected components in partially interdependent networks. 

In this work we define a mutually connected component (MCC), in a multiplex, to be a maximal set of nodes that are connected on all the layers they participate on. In other words, for interdependent networks: for any layer, all the nodes of the MCC that are present on the given layer must form a connected cluster on this layer. This definition is a straightforward interpretation of the widely accepted recursive definition.
Note that according to this definition it is possible for MCCs to overlap: nodes that only participate on one layer may belong to more than one MCC (see the example in Fig. \ref{fig_diagram_partial} (b). However, interdependent nodes (nodes that are present on all layers) can only be members of one MCC. Therefore 
 interdependent nodes form the disjunct ``cores'' of MCCs in partially interdependent networks.
In actual simulations one needs to identify all MCCs and choose the largest of them, to identify the LMCC.

We compared the results of our Effective Multiplex Degree method against other methods as a function of $f$, Fig. \ref{fig_partial}.
Note that the \texttt{EMD} algorithm naturally generalizes to partially interdependent networks.
When $f=0$, in the absence of any interdependence, naturally the single layer methods \texttt{CoreHDs} and $\texttt{CIp}_\texttt{S}$ perform best. In this case \texttt{EMD} simply returns the node degree.
As $f$ increases, the performance of \texttt{EMD} steadily improves relative to the single layer methods. Beyond approximately $f=1/2$ it outperforms all other methods except \texttt{CoreHD1}. The relatively strong performance of \texttt{CoreHD1} for moderately large values of $f$ is intriguing, but we do not know of a simple explanation. Our \texttt{EMD} method outperforms all other studied methods beyond the point where approximately one third of the nodes are non-interdependent ($\tilde{f}=2/3$, $f = 0.8$). In this case this point coincides with the value of $f$ below which the non-interpendent nodes form a giant cluster by themselves.

\section{Conclusions}

The problem of finding the minimal damaging set of even a single layer network is already a difficult problem, though recent effort has produced simple algorithms that produce highly effective approximations of the minimal set. Here our aim is to extend this effort to interdependent and multiplex networks. Due to the non-local definition of mutually connected clusters (the commonly used generalization of connected cluster to multi-layer systems) -- there is no local pruning algorithm that can give the LMCC -- this task is even more difficult. We show that reasonable generalizations of single network methods, while more effective than random attacks, are far from optimal. Rather surprisingly, direct attacks based solely on node degrees are more effective.

We have developed a simple heuristic algorithm that exploits the heterogeneity between nodes' degrees in different layers, and the vulnerability of nodes in the neighborhood to give a damaging set significantly smaller than that found by any other computationally non-intensive methods, and approaching the damaging set size found by computationally costly simulated annealing. The advantage of our method is that, due to its low computational overhead, it may be applied to very large systems, where Monte-Carlo methods like simulated annealing would be prohibitively expensive (in time and computational power) to carry out.

Our method naturally generalizes to partially interdependent networks, and remains more effective than other methods so long as interdependent nodes remain in a significant majority. When there is a significant difference in density between two layers, the connectivity of the system is controlled by the less dense layer, and single layer methods again become the most effective.
 Our algorithm also naturally extends to more general formulations of dependency within networks, for example the case of networks of ``connectivity" links and ``dependency" links \cite{parshani2011critical}. 
 A study of the effectiveness of Effective Multiplex Degree in such networks is left for future work.

\begin{acknowledgments} 
This work was supported by National Funds through FCT - Portuguese Foundation for Science and Technology, I.P. project IF/00726/2015. GT was partially supported by FCT project BPD-35/I3N/SET2016.
\end{acknowledgments}

\appendix


\section{Computation of Effective Multiplex Degree algorithm}\label{CONV}

In this appendix we discuss the computation of the Effective Multiplex Degrees (EMDs) on an arbitrary multiplex network. We show that this algorithm is computationally efficient, essentially preserving the complexity of existing popular strategies, while providing considerably better results.

The EMD values are computed iteratively according to
\begin{equation}\label{w_i_itn}
w_i^{(t+1)} = \sum_{l=1}^M \sum_{j \in \mathcal{N}^{(l)}_i} \frac{1}{M_j}\frac{w_j^{(t)}}{q_j^{(l)}},
\end{equation}
where $\mathcal{N}^{(l)}_i$ is the set of neighbours of node $i$ on layer $l$ and $q_j^{(l)}$ is the degree of node $j$ on layer $l$. 
Note that Eq. (\ref{w_i_itn}) can also be used for partially interdependent networks. In this case some of the $\mathcal{N}^{(l)}_i$ will be empty sets and some of the $q_i^{(l)}$ will be 0. 
We set the initial values $w_i^{(t=0)}$ to the sum of the degrees of nodes $i$ over all the layers (see Eq. (4)). 
In most cases this is a good initial guess, especially in partially interdependent networks for small values of $f$ (fraction of interdependent nodes). For $f=0$ this initial guess coincides with the actual solution. In Eq. (\ref{w_i_itn}) we only consider nodes and links inside the LMCC.
In this case the matrix involved in Eq. (\ref{w_i_itn}) - Eq. (\ref{R}) in the main text - is a left stochastic matrix (i.e., has a largest eigenvalue $1$), the weights $w_i$ converge to the components of the principal right eigenvector.

The number of operations required in computing one iteration of Eq. (\ref{w_i_itn}) is $\sim L$, the number of links in the network, as we have to sum over all the neighbours on all the layers for every node. The overall time complexity of our algorithm is therefore
\begin{equation}
  T \sim n_{iter} L,
  \label{eq: corr-fun1}
\end{equation}
where $n_{iter}$ is the number of iterations to convergence. Assuming $n_{iter}$ to be a constant number, the \texttt{EMD} algorithm runs in time linear in system size, achieving essentially the same efficiency as that of \texttt{DegSum} and \texttt{CoreHD}. 
(Of course, considering the complete problem of destroying the largest mutually connected component, we must repeat the identification of the LMCC and the node selection step for every node in the damaging set. This applies to all of the adaptive strategies considered.)

As shown in Fig. \ref {fig_w_convergence} (b), typically only a few (3-5) iterations of Eq. (\ref{w_i_itn}) are needed to achieve practically the same result as for the fully converged solutions. (Our criterion for the solutions to have ``fully converged'' is that the largest relative difference of all the $w_i$ values in an iteration be less than $10^{-7}$.) This proved to be the case for a range of $f$ values and different types of network architectures, so indeed it seems reasonable to consider $n_{iter}$ to be a small constant.

\subsection*{Method to identify the components}

To identify MCCs in a partially interdependent network, one can do the following.
First identify all the connected components on layer 1, and call these ``candidate components''. (These candidate components may contain interdependent nodes and free nodes on layer 1.) Then consider the next layer, check if the interdependent nodes of the existing candidate components form connected components on layer 2 (also considering free nodes on layer 2). If any of the interdependent nodes of existing candidate components are disconnected on layer 2, then subdivide them into connected components (connected on layer 2, considering also free nodes on layer 2) and mark them as new candidate components. Repeating this procedure on every layer, make as many subdivisions as necessary until the point when a full iteration (checking all the layers sequentially) can be done without further subdivisions. At this point the interdependent nodes in the existing candidate components are the "cores" of MCCs in the network. As mentioned earlier, MCCs in partially interdependent networks may overlap, but the interdependent cores of MCCs are disjunct. 

To find all the members of an MCC that a given core belongs to, simply search for all the nodes reachable from the interdependent members of each core, on each layer. This step may involve the addition of a large number of free nodes to the MCC. Because MCCs can overlap - and the overlap may be large - it is generally not efficient to store all the members of all MCCs. A good strategy is to just store the interdependent cores of MCCs (these cannot overlap), then the actual MCC that a given core belongs to can be quickly reproduced in a simple search on all the layers.

Because MCCs can overlap in partially interdependent networks, their sizes do not obey any obvious normalization condition, and their size distribution, to our knowledge, has not yet been investigated. Such studies are outside the scope of this paper, where we only concentrated on identifying the largest of the MCCs, i.e., the one that has the most members (free or interdependent).

\section{Generalisation of existing strategies to multiplex networks}\label{strategies}

In this appendix we discuss the multiplex generalizations of some of the most efficient single-layer damage schemes. Of the great number of possibilities we concentrate on some simple and meaningful generalizations that formed the basis for our comparison of the \texttt{EMD} strategy with other efficient algorithms. All of the algorithms considered in this work - including \texttt{EMD} - are ``adaptive'', meaning that after removing a node, all of the weights of the remaining nodes are recalculated.

\subsection*{Highest Degree}

The multiplex generalization of the Highest Degree strategy is simple and straightforward. Each node has a well-defined degree on each of the layers in which it participates. We considered two ways of combining these degrees. We define the \texttt{DegSum} score of node $i$ as the sum of the degrees of node $i$ over all the layers in which it participates. Similarly, let the $\texttt{DegProd}$ score be the product of the degrees of node $i$ over all the layers in which it participates. These two strategies provide similar results, which are - somewhat surprisingly - significantly better than results from the multiplex versions of the more sophisticated \texttt{CoreHD} and \texttt{CIp} algorithms.

\subsection*{\texttt{CoreHD}}

Many possibilities arise when aiming to generalize the \texttt{CoreHD} algorithm to multiplex networks. Firstly, one may consider the intersection of the 2-cores of the separate layers, or the union of them. We have tried variations of both and found that strategies where only the intersection of the 2-cores was considered always performed noticably worse than methods dealing with the union of the 2-cores.
It appears that considering only the intersection is not a good strategy, because a node that is the member of a 2-core only on one layer may still play a more important role - by virtue of this one layer - than some other less connected nodes that happen to be in the intersection of the 2-cores.
Due to these considerations we focus on generalizations of \texttt{CoreHD} that attack the union of the 2-cores of the separate layers. 

A second question that arises is how to combine the \texttt{CoreHD} scores of different layers. We considered both a ``sum'' and a ``product'' version, $\texttt{CoreHD}_\texttt{S}$ and $\texttt{CoreHD}_\texttt{P}$ respectively. For a given node $i$ these methods simply compute the sum or the product of the \texttt{CoreHD} scores for each layer of node $i$. If node $i$ is in the intersection of the 2-cores, then it has a well-defined \texttt{CoreHD} score ($\geq2$) on each layer. If on any of the layers node $i$ does not belong to the 2-core, we assign it a score $0$ for the given layer in the case of $\texttt{CoreHD}_\texttt{S}$ and $1$ in the case of $\texttt{CoreHD}_\texttt{P}$. This scheme applies also to partially interdependent networks: a node certainly does not belong to the 2-core of a layer in which it does not participate, and so it is assigned a score as explained above.
When none of the layers contain a 2-core, i.e., all layers are trees, we proceed with the $\texttt{DegSum}$ algorithm to break the MCCs up into sub-extensive sizes. This generally only takes a small number of steps and the actual method of tree-breaking does not play a relevant role.
For the majority of networks considered $\texttt{CoreHD}_\texttt{S}$ proved to be the better strategy, therefore this was chosen in the main text for comparison with other methods. (In most cases $\texttt{CoreHD}_\texttt{P}$ produced very similar, slightly worse results.)

As pointed out in the main text, when one of the layers in the multiplex network is much sparser than the others, then this layer may essentially determine connectivity in the network. Destroying the LMCC in this case corresponds closely to destroying the largest component in the weakest (most sparse) layer, therefore single layer attack strategies can be expected to work very well. We considered an algorithm - \texttt{CoreHD1} - where, in every step, we find the layer on which the LMCC has the smallest mean degree. Then we assign weights to nodes using the original \texttt{CoreHD} method on this layer. As can be seen in Fig. 4 of the main text, \texttt{CoreHD1} is a very good candidate for multiplex networks with highly different layer densities, but also performs well in symmetric networks with a moderate fraction of interdependent nodes.

\subsection*{Collective Influence Propagation}

We considered a multiplex generalization of the Collective Influence Propagation (\texttt{CIp}) algorithm \cite{morone2016collective}. For simplex networks this algorithm was shown to work better than the previous Collective Influence algorithm, so we chose only to focus on a generalization of \texttt{CIp}. This algorithm assigns a number to all nodes in a single layer: for nodes in the 2-core the score is positive, and for nodes outside the 2-core the score is $0$. Again we have a choice of how to combine the scores of nodes on different layers. Unlike in the case of simplex networks, where only the relative value of scores matters, here the normalization of the \texttt{CIp} scores becomes relevant, because the scores of nodes on different layers must be comparable in a meaningful way. (Note that this issue does not arise in the case of \texttt{CoreHD}, where the scores are simply the degrees - inside the 2-core - of nodes in the 2-core.) An intuitively appropriate normalization scheme is to have the \texttt{CIp} scores on each layer sum to $N_m$, the number of nodes on layer $m$. Then, we define the values $\texttt{CIp}_\texttt{S}(i)$ for each node $i$ as the sum of the (correctly normalized) $\texttt{CIp}(i)$ values over all the layers in which node $i$ participates. 

We did not consider the ``product'' version of the \texttt{CIp} algorithm, because in this case the following problem arises. Consider two nodes on a given layer that have the same $\texttt{CIp}$ values (on this layer). Assume that one of the nodes is present only on this layer, but the other one participates on all other layers. If the latter (interdependent) node has $\texttt{CIp}$ values less than $1$ on the other layers, it will receive a score that is less than the score of the first (free) node, if the scores of different layers are combined as a product. This is rather counterintuitive. There are certainly ways of circumventing this problem, but such methods would look rather contrived and arbitrary, therefore we chose to consider only the ``sum'' version, $\texttt{CIp}_\texttt{S}$.

\section{Simulated Annealing}\label{SA}

As a benchmark, against which all other strategies can be compared, we used a Simulated Annealing (\texttt{SA}) algorithm to identify a close upper bound on the size of the minimal damaging set that makes the LMCC sub-extensive. Our \texttt{SA} method employs a standard Metropolis Monte Carlo algorithm as follows. We consider the configuration space where all nodes in the network (free or interdependent) can be in one of two states: \textit{present} or \textit{removed}. We exclude all configurations where there is an extensive LMCC in the network, when only nodes present are considered. (Our criterion for ``extensive'' is to have a size larger than $\sqrt{N}$, the square root of the network size.) We define the energy of a configuration simply as the number $R$ of nodes removed. We simulate this system in thermal equilibrium at temperature $T$ by applying the following steps repeatedly:

\begin{enumerate}

\item

We choose a node in the network uniformly at random and make the trial change of switching its state to the opposite. (If it was \textit{removed}, switch to \textit{present} and if it was \textit{present}, switch to \textit{removed}.)

\item

Find the size of the LMCC in the network and if it is extensive ($ > \sqrt{N} $), reverse the trial change and go to Step 1.

\item

If the LMCC after the trial change is still sub-extensive, accept the trial change with probability $p = min(1, e^{-\Delta R/T})$ and go to Step 1.
 
\end{enumerate}

As initial condition, we set all nodes to \textit{removed}, i.e., $R = N$. The initial temperature for the simulations was $T_{max} = 1$ and in every step it was reduced by $\delta T$ until $T_{min} = 0.01$ was reached. $\delta T = 10^{-7}$ was used for networks of size $N = 10^4$, and in a reasonable running time results were all considerably better than those provided by faster attack strategies.
For network size $N=10^5$, it was necessary to increase $\delta T$ to $10^{-6}$ in order to maintain a feasible running time.
As a result, \texttt{SA} already performed much worse than some of the best heuristic strategies.

Simulated annealing is a widely used technique that is assumed to approach the optimal solution when given sufficient time to thoroughly probe the configuration space. It is therefore a good choice as a benchmark for comparisons, but it is prohibitively slow for large system sizes, so can't be considered a viable candidate for practical purposes. Also, being a purely stochastic method, it does not give us any insight as to the structure of a good damage set.

\bibliography{targetted_multiplex}

\end{document}